\documentclass[compsoc,conference,a4paper,10pt,times]{IEEEtran}
\IEEEoverridecommandlockouts
\usepackage{cite}
\usepackage{amsmath,amssymb,amsfonts}
\usepackage{algorithmic}
\usepackage{graphicx}
\usepackage{textcomp}
\usepackage{bmpsize}
\usepackage{xcolor}
\usepackage{lipsum}
\usepackage{float}
\usepackage{bbm}
\usepackage{dsfont}
\usepackage{tcolorbox,mathrsfs}
\usepackage{xspace}
\usepackage{booktabs}
\usepackage{subcaption}
\usepackage{natbib}
\usepackage{url}
\usepackage[colorlinks=true,urlcolor=black]{hyperref}
\def\BibTeX{{\rm B\kern-.05em{\sc i\kern-.025em b}\kern-.08em
    T\kern-.1667em\lower.7ex\hbox{E}\kern-.125emX}}

\usepackage[textwidth=1cm,textsize=tiny]{todonotes}

\begin{document}

\renewcommand{\equationautorefname}{Eq.}
\renewcommand{\figureautorefname}{Fig.}
\renewcommand{\tableautorefname}{Tab.}
\renewcommand{\sectionautorefname}{Sec.}
\renewcommand{\subsectionautorefname}{Sec.}
\def\equationautorefname#1#2\null{%
  Eq.#1(#2\null)%
}

\newcommand{\mypar}[1]{\noindent\textbf{#1}}
\newcommand{\vct}[1]{\boldsymbol{#1}}
\newcommand{\takehome}[2]{
\begin{tcolorbox}[colback=blue!35, colframe = black, boxrule=0.5pt]
\textbf{Take-home Message #1:} #2
\end{tcolorbox}
}

\newcommand{\svm}{SVM\xspace}
\newcommand{\gbt}{GBDT\xspace}
\newcommand{\svmyara}{SVM\textsubscript{f}\xspace}
\newcommand{\gbtyara}{GBDT\textsubscript{f}\xspace}

\title{Demystifying the Role of Rule-based Detection\\
in AI Systems for Windows Malware Detection}

\author{
\IEEEauthorblockN{
Andrea Ponte\IEEEauthorrefmark{1},
Luca Demetrio\IEEEauthorrefmark{1},
Luca Oneto\IEEEauthorrefmark{1},
Ivan Tesfai Ogbu\IEEEauthorrefmark{3},
Battista Biggio\IEEEauthorrefmark{2},
Fabio Roli\IEEEauthorrefmark{1}\IEEEauthorrefmark{2}
}\\
\IEEEauthorblockA{\IEEEauthorrefmark{1}University of Genova, Genova, Italy\\
andrea.ponte@edu.unige.it,\\
\{luca.demetrio, luca.oneto, fabio.roli\}@unige.it}
\IEEEauthorblockA{\IEEEauthorrefmark{2}University of Cagliari, Cagliari, Italy\\
battista.biggio@unica.it}
\IEEEauthorblockA{\IEEEauthorrefmark{3}RINA Consulting S.p.A., Genova, Italy\\
ivan.tesfai@rina.org}
}

\maketitle

\begin{abstract}
Malware detection increasingly relies on AI systems that integrate signature-based detection with machine learning.
However, these components are typically developed and combined in isolation, missing opportunities to reduce data complexity and strengthen defenses against adversarial EXEmples, carefully crafted programs designed to evade detection.
Hence, in this work we investigate the influence that signature-based detection exerts on model training, when they are included inside the training pipeline.
Specifically, we compare models trained on a comprehensive dataset with an AI system whose machine learning component is trained solely on samples not already flagged by signatures.
Our results demonstrate improved robustness to both adversarial EXEmples and temporal data drift, although this comes at the cost of a fixed lower bound on false positives, driven by suboptimal rule selection.
We conclude by discussing these limitations and outlining how future research could extend AI-based malware detection to include dynamic analysis, thereby further enhancing system resilience.
\end{abstract}


\section{Introduction}

To increase the likelihood of detecting the always-evolving variants of malware, malware detectors are enriched with several layers of detection, both relying on pattern matching and machine learning (ML) components.
We name such combination as \emph{AI systems}, aligned with the latest directives of the European Union through the EU AI Act~\cite{aiact}.
This is consistently reported as the standard by companies that sell antivirus (AV) programs~\cite{eset,avira,windefender,macos}, while academia is recently starting to investigate the problem of composing sequential layers of detection modules~\cite{ponte2025slifer}.
Regardless of their provenience, either from industry or academia, the first layer of detection is achieved through pattern-matching with YARA rules.\footnote{\url{https://github.com/VirusTotal/yara}}
Manually crafted by domain experts, YARA rules contain both specific patterns of byte associated with malicious samples, and conditions that must be met to trigger detection.
However, even if many samples are already stopped by available rules, malware detectors are trained on all available data, and deployed alongside YARA rules.
While intuitive, this methodology might not be optimal, since each layer of detection alters the distribution of data fed in input to the subsequent detection module.
As a result, models are trained on samples they will never see at test time, complicating not only the training process due to the massive amount of samples used to create these detectors, but also potentially including spurious correlations exploited by attackers with \emph{adversarial EXEmples} -- carefully-manipulated Windows malware that try to evade ML detection~\cite{demetrio2019explaining, demetrio2021adversarial}.
\begin{figure}[t!]
    \centering
    \includegraphics[width=\linewidth]{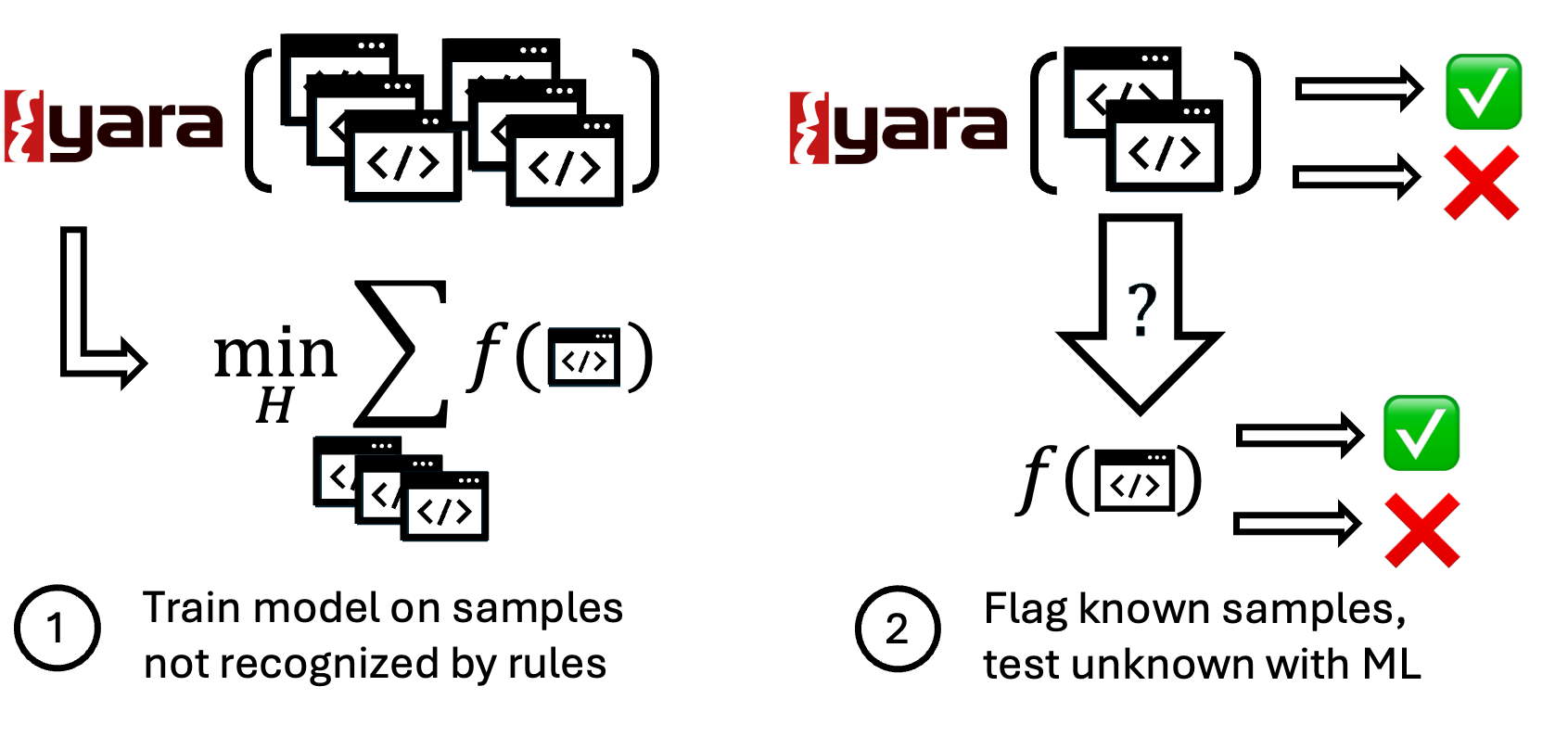}
    \caption{AI System for malware detection, which combines signature-based detection to ML, both during (1) training, by removing samples that trigger YARA rules, and (2) test time, by using ML only on unknown samples.}
    \label{fig:example}
\end{figure}

Hence, in this work we perform preliminary steps towards an empirical understanding of the role of YARA detection layer at training and test time, as shown in \autoref{fig:example}.
As far as we know, we are the first to propose such an analysis, where we exclude from training data \emph{all} those samples that are either matched by at least one YARA rule, or are contained inside an allowlist composed utilities harvested from fresh installation of Windows.
We empirically quantify how such pre-filtering influences the learning process of the AI system we build, showing that this methodology matches and exceeds the performances of models trained on all the available samples at low FPR (1\%), also when exposed to never-seen future data possibly affected by concept drift.
Also, thanks to the presence of YARA rules, the AI system we build exhibits higher robustness to adversarial EXEmples rather then regularly-trained models, since the decision function is harder to explore by the attack.
Similar to previous work~\cite{ponte2025slifer}, we observe that the computation of adversarial EXEmples can introduce unforeseeable artifacts that are detected by rules when injected by the optimization process. 
However, these benefits are balanced by a fixed amount of false positives induced by rules themselves.
Even if some of them are very effective in stopping hundreds of malware samples, they can not be tuned to adapt to the training data and reduce their response to false alarms, differently from machine learning models.
We conclude our work by remarking that these are preliminary results on the performance of AI systems for malware detection, but already highlighting potential improvement in the field in terms of efficiency and robustness.

\section{Background and Related Work}
Before describing our methodology, we introduce the key concepts needed to fully understand our manuscript, comprising the types of data we are dealing with, how to detect malicious samples among them, and how attackers can evolve their technique to evade complex ML models used for malware detection.
\begin{figure}
    \centering
    \includegraphics[width=\linewidth]{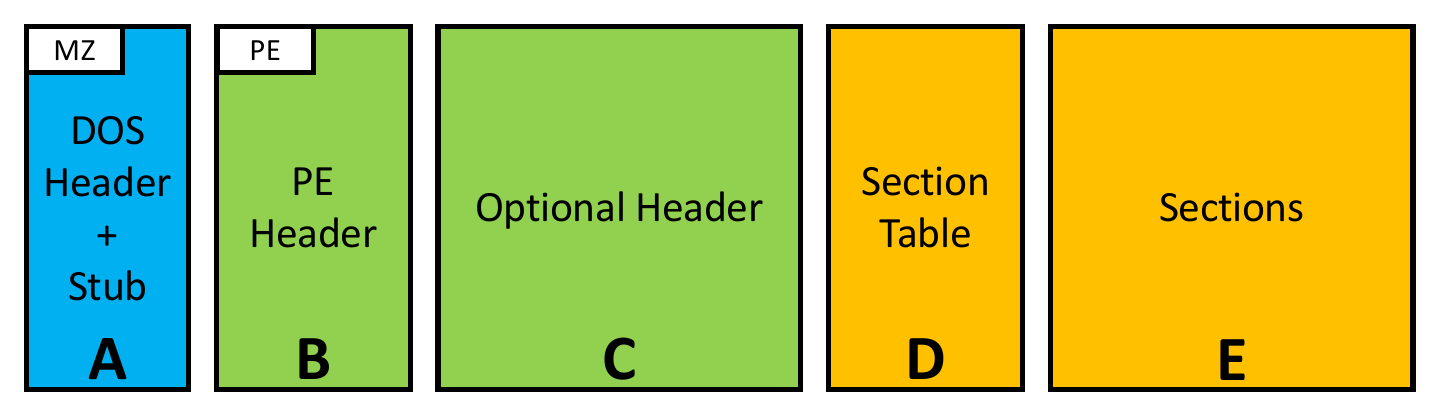}
    \caption{Depiction of the Windows PE File Format.}
    \label{fig:pe}
\end{figure}

\mypar{Windows PE File Format.} Each Windows program is stored as a file, whose structure is defined by the Portable Executable (PE) format\footnote{\url{https://learn.microsoft.com/en-us/windows/win32/debug/pe-format}}.
As depicted \autoref{fig:pe}, the format provides precise information on how to load programs in memory, and it is divided into metadata and code.
The latter, alongside other relevant information such as initialized data, are stored in multiple \emph{sections} (E in \autoref{fig:pe}), which constitute the majority of each program.

\mypar{Detection with YARA Rules.}
YARA is a pattern-matching tool used to detect known malware based on signatures. These signatures are textual descriptions containing binary patterns that help identify malicious programs. 
Each YARA rule consists of metadata that describes its purpose, followed by strings or patterns—such as hexadecimal sequences or regular expressions—that the tool searches for within files. 
Finally, each rule includes a firing condition, which determines how malicious activity is identified. 
This condition is structured as an if-then-else logic block that defines the detection algorithm. 
This mechanism is not only helpful in detecting malware samples, since conditions can be designed to identify well-known unharmful executables, such as operating systems applications and certified software installers.

\mypar{Static Malware Detection with Machine Learning.}
This type of analysis discriminates legitimate and malicious programs by analyzing the sole structure of executables without running them, and train machine learning models on top of these activities.
The latter is achieved by training models either on features, which are measurements computed on data through dedicated algorithms that leverage domain knowledge~\cite{anderson2018ember, saxe2015deep}; or on raw bytes~\cite{raff2018malware} often treated as gray-scale images~\cite{nataraj2011malware}; or on a combination of the two mentioned approaches~\cite{liu2017automatic}.
While promising, static analysis might be circumvented by obfuscation since malware samples can manifest malicious behaviors at runtime that cannot be easily inferred without execution. 
Also, complex feature extraction algorithms might crash on specific samples, requiring flexibility in the analysis process.
To overcome these limitations, static malware detectors can be improved by either applying redundant and multiple controls~\cite{ponte2025slifer} also involving signature-based detection, or merging it with \emph{dynamic} analysis~\cite{trizna2022quo}, which also captures the runtime behavior of samples.
However, as far as we know, none of these work on either single- or multi-layer detectors considered the influence that each layer imposes on the next.
In fact, no research work has investigated the effect of the change of data distribution between layers, and how this affects the final performances in production.

\mypar{Adversarial EXEmples.}
As previously shown in literature, Windows malware detectors can be ineffective against  \emph{Adversarial EXEmples}, carefully-manipulated programs that evade detection without altering their original functionality~\cite{demetrio2019explaining, demetrio2021functionality, demetrio2021adversarial}.
These samples are crafted through optimization algorithms that only rely on the answer of the target model to evade, manipulating the samples through content-injection procedures.
To improve the likelihood of success, there are techniques like GAMMA~\cite{demetrio2021functionality} that fools the target model by injecting into malware sample content extracted from legitimate programs, without altering the flow of execution.

\section{Experimental Analysis}\label{sec:experimental_analysis}
Our methodology revolves around removing samples from the training data that are already detected by pre-defined YARA rules, and then we train a machine learning model only on the undetected ones.
Hence, we start by describing the setup (\autoref{sec:setup}), and how the select dataset is filtered by YARA rules (\autoref{sec:filtering}).
We empirically analyze the effect of such a filtering on the performances at test time in terms of accuracy and false positives (\autoref{sec:performance}), and its robustness against attacks (\autoref{sec:robustness}).

\subsection{Experimental Setup}
\label{sec:setup}
In this subsection, we describe how we setup the experiments and which data we leverage to train and test the machine learning models.

\mypar{Dataset.}
We use the Speakeasy dataset~\cite{trizna2022quo} as the main source of malware and goodware samples. 
It comprises several malware families as reported in~\cite{trizna2022quo} and is divided into training and test data. 
This dataset is unbalanced, and it counts 71506 malware samples and 26059 goodware samples.\footnote{These numbers are slightly different from the ones reported by authors of the dataset, as we removed some duplicates we found before conducting our experiments.}
To deal with such an imbalance, we increase the number of benign samples in the training set by complementing it (i) with 9864 goodware samples collected from Chocolatey;\footnote{\url{https://community.chocolatey.org/}} and (ii) with system files stored in \textit{sys32} and \textit{syswow64} of fresh installation of Windows 8.1, 10 and 11, for a total of 2648 goodware. 
Inside Chocolatey goodware samples, we found 54 Windows system files already present in our fresh installations, that we discard to avoid repetitions.
We also add 5000 malware samples and 3446 goodware samples used in~\cite{demetrio2021functionality} and \cite{ponte2025slifer}.
The Speakeasy dataset is characterized by a temporal shift between the training (collected in January 2022) and the test data (collected in April 2022), which is is useful for testing the performance of the models in an evolving scenario. 
Hence, we name our corpus of training data of Speakeasy and all the additions we have mentioned as Present Data (counting 76506 malware and 41952 goodware samples), and the Speakeasy test data (17500 malware and 10000 goodware) as Future Data. 
However, the goodware samples included in the training belong to the same time period of the ones in the test, due to the inclusion of the goodware samples from Chocolaty to improve the balance between classes.
We are aware that such a setting could lead to \textit{data snooping}~\cite{arp2022and}, but such injection of goodware from other sources addresses the problem of the class imbalance.
We measure the performances of models on Present Data building a test set composed of 10\% of the total amount of samples (7685 malware and 4162 goodware).
For brevity we indicate the Present test set as $\mathbf{T_p}$ and the Future test set as $\mathbf{T_f}$.
For each file of the dataset, we compute the EMBER features~\cite{anderson2018ember}, which represents a golden standard in static analysis feature extraction. 
The feature extraction failed for 4 samples, which we excluded from the datasets.

\mypar{YARA Rules.}
We collect YARA rules for our blocklist from several public GitHub repositories\footnote{\url{https://github.com/bartblaze/Yara-rules/tree/master/rules}}$^,$\footnote{\url{https://github.com/elastic/protections-artifacts/tree/main/yara/rules}}$^,$\footnote{\url{https://github.com/malpedia/signator-rules/tree/main/rules}}$^,$\footnote{\url{https://github.com/Neo23x0/signature-base/tree/master/yara}}$^,$\footnote{\url{https://github.com/Yara-Rules/rules/tree/master/malware}}, collecting 2.7K rules. 
As regards the allowlist, we created a rule containing the hashes of the Windows system files inside our dataset, that checks if the input is a well-known legitimate system program.

\mypar{Selected Architectures.}
We select two ML models for our experiments: a Support Vector Machine~\cite{cristianini2008support, shalev2014understanding} (\svm) with Gaussian kernel implemented in SciKit-Learn~\cite{scikit-learn} and a Gradient Boosted Decision Tree (\gbt)~\cite{friedman2001greedy} implemented with the XGBoost library~\cite{chen2016xgboost}.
The best hyper-parameters $\gamma$ and $C$ for the \svm are chosen after an extensive model selection~\cite{oneto2020model} choosing $\gamma\in10^{\{-6.0, -5.5, ..., 4.0\}}$ and $C\in10^{\{-6.0, -5.5, ..., 4.0\}}$, trying 30 values spaced evenly on a log scale, for both parameters. 
We perform a grid-search cross-validation with 10 folds, and we use a subset of our large training with 30K samples for each class. 
The resulting values for the hyper-parameters are used for both the regularly-trained models and the AI system trained on filtered data.
For \gbt we set 1000 trees and $\eta=0.1$, the subsample ratio of columns when constructing each tree to $0.8$, while keeping default values for all the other parameters.
We produce models trained only on data that do not trigger any YARA rule denoted as \svmyara and \gbtyara, and we compare them with models trained on all the data we possess (noted as \svm and \gbt).

\mypar{Setup of Adversarial Attacks.}
The robustness evaluation of all the models we train is performed with GAMMA, that leverages the injection of new sections harvested from legitimate programs~\cite{demetrio2021functionality} to manipulate malware.
The attacks are conducted on a subset of the test set $\mathbf{T_p}$ and on a subset of $\mathbf{T_f}$, randomly selecting 100 malware samples from each family in the Speakeasy dataset (resulting in a total of 700 samples gathered from both sets). 
Since we are interested in the effect imposed by rules filtering, which considers different data distributions at training time, we setup GAMMA to harvest content from two sources: (i) legitimate Windows utility programs extracted from a fresh installation of Windows 10; and (ii) samples scraped from Chocolatey\footnote{\url{docs.chocolatey.org}} which is a well-known package manager for Windows.
In both cases, we test attacks that can query the target at maximum 200 times, with increasing forces, allowing GAMMA to select content harvested from 10, 20, 30, and 50 sections among these legitimate programs.
This procedure leads to 32 attack configurations for both models.
For the SVM models, we fix the regularization term $\lambda = 10^{-5}$ and the number of queries to 200 for all configurations. 
For the \gbt models we fix the number of queries to 500, and we set $\lambda = 10^{-7}$ when using 10 to 20 sections extracted from Windows goodware, while $\lambda = 10^{-8}$ for 30 and 50 sections from the same source. Instead, when using Chocolatey PEs which are bigger in size, we set $\lambda = 10^{-6}$ for 10 and 20 sections attacks and $\lambda = 10^{-7}$ for 30 and 50 sections attacks.

\mypar{Evaluation Metrics.}
For performance and temporal analysis, we compute ROC curves, which show the True Positive Rate (TPR) and False Positive Rates (FPR) varying the decision threshold of the model. 
Regarding the robustness to adversarial EXEmples, we compute the Detection Rate (which is the TPR) at 1/\% FPR as the manipulation size increases. 

\mypar{Hardware.}
To compute all our experiments, we leverage a workstation equipped with an Intel® Xeon(R) Gold 5420, two Nvidia L40 GPUs, and 540 GB of RAM. 

\begin{table}[t]
\centering
\begin{tabular}{@{}cccccc@{}}
\toprule
\multicolumn{2}{c}{$\mathbf{D_t}$}     & \multicolumn{2}{c}{$\mathbf{T_p}$}    & \multicolumn{2}{c}{$\mathbf{T_f}$}                                 \\ \midrule
\textbf{TPR} & \textbf{FPR} & \textbf{TPR} & \textbf{FPR} & \textbf{TPR}          & \textbf{FPR} \\ \midrule
     0.24       &     0.0094        &       0.23       &      0.0096       & 0.34        &             0.0089           \\ \bottomrule
\end{tabular}
\caption{YARA performance (represented by TPR and FPR) on the training set $\mathbf{D_t}$, the present test set $\mathbf{T_p}$ and the future test set $\mathbf{T_f}$.}
\label{table:yara_performance}
\end{table}

\begin{figure*}[t]
    \centering
    \begin{subfigure}{0.49\linewidth}
        \centering
        \includegraphics[width=\linewidth]{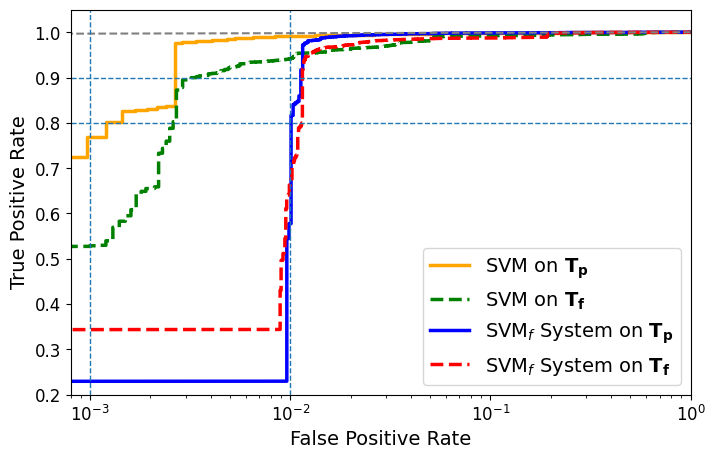}
        \caption{}
        \label{fig:svm_roc}
    \end{subfigure}
    \hfill
    \begin{subfigure}{0.49\linewidth}
        \centering
        \includegraphics[width=\linewidth]{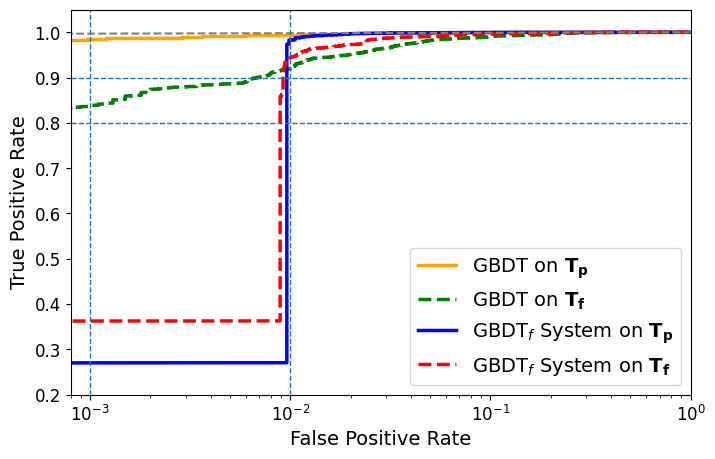}
        \caption{}
        \label{fig:xgb_roc}
    \end{subfigure}
    \caption{ROC curves of all models and systems considered. We report \svm and \svmyara System performances in \autoref{fig:svm_roc}, while \gbt and \gbtyara System performances are illustrated in \autoref{fig:xgb_roc}. Models trained on all data and AI systems that integrate YARA rules are tested with $\mathbf{T_p}$ (solid lines) and $\mathbf{T_f}$ (dashed lines).}
    \label{fig:roc_comparison}
\end{figure*}

\subsection{Data Filtering}
\label{sec:filtering}
We train our models in two different configurations: (i) by filtering our training dataset with our YARA blocklist and allowlist thus training ML models only on data not recognized by signatures; and (ii) by using all the training data available, without filtering. 
Before we proceed towards training models though, we need to quantify the predictive capabilities in terms of TPR and FPR of the rules we have collected, and we present this analysis in  \autoref{table:yara_performance}.
The allowlist, which targets utilities of Windows, filters all the 2648 system files in the dataset as expected, while the blocklist removes 16329 malicious samples from the training data.
At test time, the allowlist blocks 249 benign samples belonging to $\mathbf{T_p}$ and 0 samples belonging to $\mathbf{T_f}$. Instead, the blocklist blocks 1765 malicious samples of $\mathbf{T_p}$ and 6016 samples of $\mathbf{T_f}$. In other terms, the blocklist detect 24\% of malware inside the Present Data ($\mathbf{D_t}$ and $\mathbf{T_p}$) and 34\% of malware in Future Data ($\mathbf{T_f}$).
For all datasets, we note that a non-negligible amount of FPR $\approx 10^{-2}$.
By evaluating those through the responses of VirusTotal\footnote{\url{https://www.virustotal.com/gui/home/search}}, we discover that they are caused by (i) \textit{grayware programs} contained in the data we have described in \autoref{sec:setup}, (ii) samples flagged by less than five AVs, or legitimate Python installers.

\subsection{Performance and Temporal Analysis}
\label{sec:performance}
We analyze the performance of the models we have trained, using both $\mathbf{T_p}$ and $\mathbf{T_f}$.
We report in \autoref{fig:roc_comparison} the ROC curves of \svm and \gbt trained on all data, and we evaluate the same curves on corresponding AI systems trained on data filtered by the YARA rules.
%
Regarding regularly-trained models, both \svm and \gbt achieve high TPR on $\mathbf{T_p}$ at $1\%$ FPR, with \svm experiencing a consistent drop at extremely low FPR ($0.1\%$). 
However, both models manifest a decay in performance when evaluated on $\mathbf{T_f}$ (green dashed line in \autoref{fig:roc_comparison}).
While \gbt confirms its capability at extremely low FPR, \svm maintains good performance only at 1\%.
As regards AI systems, ROC curves highlights the contribution of YARA rules, which guarantee a fixed TPR.
However, they also force a fixed FPR, shown by the horizontal ROC until they reach almost $1\%$ FPR. 
In fact, YARA rules report a fixed FPR of $0.96\%$ on $\mathbf{T_p}$ and $0.89\%$ on $\mathbf{T_f}$ as reported in \autoref{table:yara_performance}. 
Beyond those points, the actual performance of \svmyara and \gbtyara becomes visible, showing similar or even improved results compared to models trained on the entire dataset. 
We see the AI system with \gbtyara outperforming \gbt at $1\%$ FPR on $\mathbf{T_f}$ while scoring almost the same on $\mathbf{T_p}$. 
The same holds for the AI system with \svmyara compared to \svm, but at a slightly higher level of FPR.
Lastly, we clarify why there is a mismatch between TPRs of AI Systems at very low FPR (left side of \autoref{fig:svm_roc} and \autoref{fig:xgb_roc}). 
The AI System built with \gbt achieves higher TPR than the one built with \svm, and
this difference is caused by some samples being detected by \gbt with full confidence, i.e. the output of the model is precisely 1.
Hence, while computing the ROC it is not possible to fix a threshold for which those are misclassified as benign, thus causing the discrepancy with SVM, which does not exhibit this behavior.
\takehome{1}{AI Systems can be trained on fewer data, matching the performance of models trained on all data. However, such performances are limited by the fixed FPR of YARA rules.}

\subsection{Robustness Analysis}
\label{sec:robustness}
\begin{figure*}[]
    \centering
    \begin{subfigure}{0.49\linewidth}
        \centering
        \includegraphics[width=\linewidth]{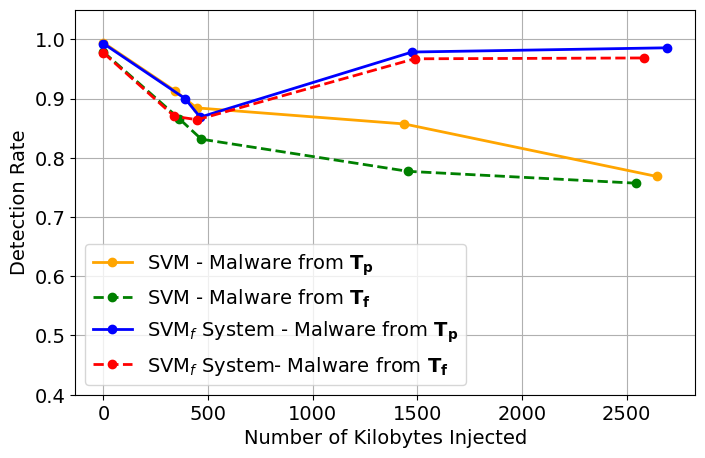}
        \caption{}
        \label{fig:svm_dr_choco}
    \end{subfigure}
    \hfill
    \begin{subfigure}{0.49\linewidth}
        \centering
        \includegraphics[width=\linewidth]{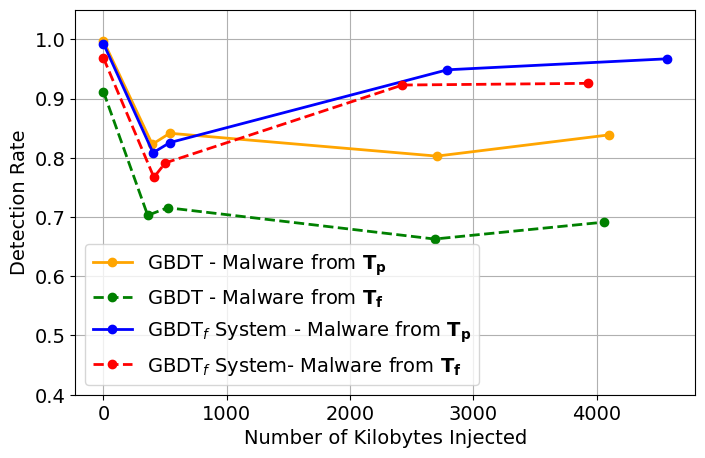}
        \caption{}
        \label{fig:xgb_dr_choco}
    \end{subfigure}
    

    \begin{subfigure}{0.49\linewidth}
        \centering
        \includegraphics[width=\linewidth]{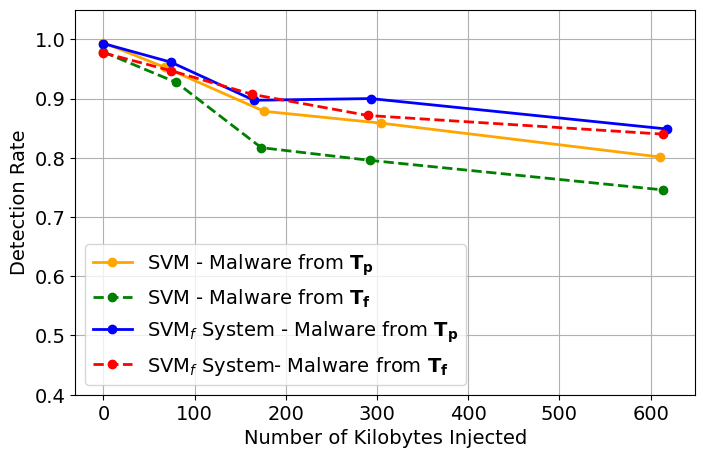}
        \caption{}
        \label{fig:svm_dr_win}
    \end{subfigure}
    \hfill
    \begin{subfigure}{0.49\linewidth}
        \centering
        \includegraphics[width=\linewidth]{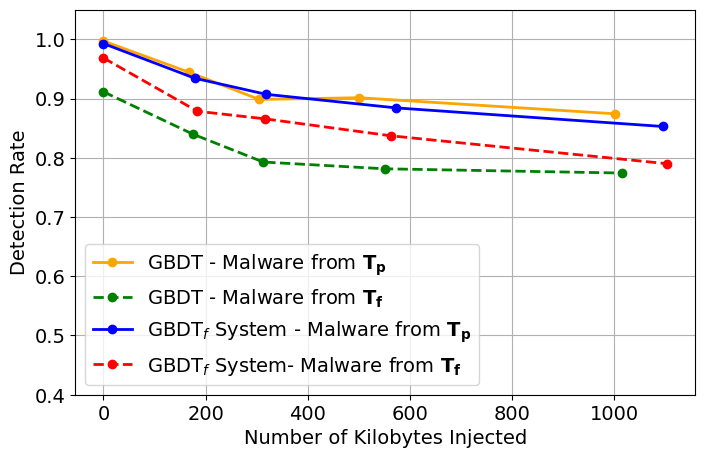}
        \caption{}
        \label{fig:xgb_dr_win}
    \end{subfigure}

    \caption{Detection rates of SVM, GBDT, \svmyara System, and \gbtyara System injecting Chocolatey (\autoref{fig:svm_dr_choco} and \autoref{fig:xgb_dr_choco}) and Windows 10 (\autoref{fig:svm_dr_win} and \autoref{fig:xgb_dr_win}) benign sections. Detection rates are reported based on the number of kilobytes injected (which varies according to \textit{sections} parameter of GAMMA) and the source of malware samples. We report detection rates of adversarial EXEmples originated from $\mathbf{T_p}$ and $\mathbf{T_f}$ malicious samples with solid lines and dashed lines respectively.}
    \label{fig:dr_combined}
\end{figure*}
The experimental analysis on robustness brings evidence of the contribution of YARA rules, coherently to what is presented by Ponte et al.~\cite{ponte2025slifer}. 
We summarize all attack configurations using programs scraped from Chocolatey as content inject in \autoref{fig:svm_dr_choco} and \autoref{fig:xgb_dr_choco}, and we report the results of attacks using Windows 10 programs in \autoref{fig:svm_dr_win} and \autoref{fig:xgb_dr_win}. 
In general, as we expect, GAMMA finds more effective manipulations starting from out-of-distribution malware samples (dashed lines in all plots), since not only the TPR of models is lower in this condition, but also the attack is likely to produce samples deviating from the original training distribution.
Also, the usage of Windows 10 benign sections results in smaller adversarial manipulations, since these programs are also smaller in size (on average) w.r.t. the Chocolatey ones, leading to lower evasion rates against \svm and \gbt, as shown in \autoref{fig:svm_dr_win} and \autoref{fig:xgb_dr_win}. 
Instead, for \svmyara and \gbtyara systems, the trend holds for the first two points (10 and 20 section manipulations), but larger manipulations show a clear divergence from \autoref{fig:svm_dr_choco} and \autoref{fig:xgb_dr_choco}.
This phenomenon is caused mainly by two facts.
The first reason is due to the inability of GAMMA of removing the patterns detected by rules from adversarial EXEmples.
Hence, when untainted malware samples trigger a certain number of YARA rules, the corresponding adversarial EXEmple also triggers the same rules.
The second reason is caused by GAMMA injecting particular strings that trigger rules, making the manipulation ineffective for evading the AI system. 
We clearly see this evidence in \autoref{fig:svm_dr_choco} and \autoref{fig:xgb_dr_choco}: when injecting 10 and 20 sections the detection rate decreases or remains approximately the same, while when injecting 30 and 50 benign sections, this trend inverts. 
Examining the phenomenon in depth, we notice that a particular benign sample is included in 30 and 50 sections attacks, and when GAMMA crafts adversarial EXEmples using that goodware sections, \texttt{Typical\_Malware\_String\_Transforms}\footnote{\url{https://github.com/Neo23x0/signature-base/blob/master/yara/gen_transformed_strings.yar}} rule is always fired.
We note that this YARA rule does not activate when tested on the original benign sample (which is a telemetry application), since it has been crafted to avoid such false positive.
However, when GAMMA injects fragments of the sample, this condition is no longer met, leading to a significant number of attacks being blocked. 
We remark that, while previous work remarked that attackers should always remove false positives from their sources during content-injection attacks~\cite{ponte2025slifer}, the artifacts we detect in this work could not be foreseen in advance.
That would have required the evaluation of each YARA rule by splitting their triggering conditions in all the single conditions chained with and/or operators, and re-evaluating them (and all their combinations) on benign samples owned by the attackers.
This would introduce a combinatorial problem into the selection of legitimate programs to use for content-injection attacks, increasing the complexity of the attack setup.
Lastly, our experimental analysis reports an unsteady trend of \gbt, which can be attributed to (i) a reduced set of considered sections (which in the original formulation of the attack is 75); and (ii) the randomness behind the genetic algorithm used to implement the attack itself.
While SVM is characterized by a smooth decision function, easier to explore, \gbt models produce piece-wise constant decision regions that requires more random sampling performed by the optimization algorithm to find meaningful directions.
Also, this can be seen in \autoref{fig:xgb_dr_choco}, since the attack reduces the detection rate with a small amount of bytes, and then keeps roughly the same results for increasing strength of the attack. 
Finally, we observe that AI systems tend to be more robust than regular machine learning models, even when sections from Windows 10 utilities are injected. As previously mentioned, this is caused by the effectiveness of the blocklist, that can still effectively detect malware even after such manipulations.

\takehome{2}{AI Systems are more robust to adversarial EXEmples due to the presence of rules.
Also, content-injection attacks might include malicious patterns that trigger YARA rules, without being extracted by false positives of signature-based detection.}

\section{Conclusions}
We now discuss the limitations of our methodology, by also highlighting the future lines of research that can take our work as a basis.

\mypar{Limitations.}
While our analysis represents a step forward in understanding the properties of AI systems, we note the following limitations.
First, the YARA rules used in this work were collected from public online repositories, and we did not filter them based on their precision on the training set.
Although this filtering might have improved the performance of the AI systems, it could also have unfairly skewed the comparison in their favor.
Moreover, our investigation would benefit from a broader model selection since we only considered one kernel for the \svm, and we did not fully explore the parameter space for the \gbt.
Furthermore, we did not include complex deep neural architectures such as the one by Harang et al.~\cite{harang2020sorel}. 
We also tested the models' robustness only through content-injection attacks implemented by injecting sections with GAMMA, and did not consider other techniques proposed in the state of the art.
In fact, other attacks might create different artifacts inside samples that might be detected by rules as well as the anomalies left by GAMMA.
Nevertheless, we believe that the results showcased in \autoref{sec:experimental_analysis} are valid, as we employed previously analyzed technologies that can be considered state of the art in this domain~\cite{anderson2018ember,demetrio2021functionality,trizna2022quo}.

\mypar{Future Work.}
While expanding our analysis to include more models and attacks is straightforward, we can further improve this investigation by exploring dynamic malware analysis, which collects the behavior of samples through execution in controlled environments (sandboxes).
In particular, we could leverage signature-based detection in this context as well, for example by deploying CAPA\footnote{\url{https://github.com/mandiant/capa/}} which identifies specific behavior patterns from sandbox reports.
Moreover, regardless of the type of analysis, we could adapt the technique proposed by Scano et al.~\cite{scano2024modsec}, which consists in training a linear model on the outputs of rules, thus weighting their answers to reduce false positives while maximizing the detection rate. Another improvement in our methodology could be using class weights at training time: tested models could benefit from this technique, possibly overcoming the problem of the unbalanced dataset.

\mypar{Final Remarks.}
In this work, we investigated AI Systems for malware detection, showing that these can match the performance of models trained on larger corpus of data while improving robustness, with a fixed amount of false alarms caused by rules themselves.
Aligned with the upcoming recommendations from the EU Commission, we believe that extensive research should be conducted on developing and evaluating AI systems, moving away from the ``in vitro'' evaluation of isolated machine learning models.
Hence, we believe that the preliminary results we collected on AI systems for malware detection could pave the way for further research, ultimately leading to the development of novel training and evaluation pipelines tailored to these technologies, which would also be compliant with emerging international guidelines and standards.

\section*{Acknowledgments}{
Andrea Ponte acknowledges the support of Rina Consulting S.p.A. for his doctoral scholarship and research work. The authors acknowledge Matous Kozak's help with data collection.
This work was partially supported by projects SERICS (PE00000014) and FAIR (PE00000013) under the NRRP MUR program funded by the EU - NGEU.
}

\bibliographystyle{unsrt}
\bibliography{biblio}

\end{document}